\documentclass{aa}  

\usepackage{graphicx}
\usepackage{txfonts}
\usepackage{hyperref}

\newcommand{\dd}[0]{\ensuremath{\mathrm{d}}}%
\newcommand{\Htwo}[0]{\ensuremath{\mathrm{H}_2}}%

\makeatletter
\renewcommand*\aa@pageof{, page \thepage{} of \pageref*{LastPage}}
\makeatother

\begin{document} 

\defcitealias{2003MNRAS.340..949B}{BEG03}
\defcitealias{2015MNRAS.449.2421S}{SBM15}

\title{Bayesian inference of three-dimensional gas maps}

\subtitle{I. Galactic CO}

\author{
P.~Mertsch\inst{1}
\and
A.~Vittino\inst{1}
}

\institute{Institute for Theoretical Physics and Cosmology (TTK), RWTH Aachen University, Sommerfeldstr. 16, 52074 Aachen, Germany\\
\email{pmertsch@physik.rwth-aachen.de}\label{inst1}
}

\abstract{
Carbon monoxide (CO) is the best tracer of Galactic molecular hydrogen (\Htwo{}). Its lowest rotational emission lines are in the radio regime and thanks to Galactic rotation emission at different distances is Doppler shifted. For a given gas flow model the observed spectra can thus be deprojected along the line of sight to infer the gas distribution. We use the CO line survey of~\citet{2001ApJ...547..792D} to reconstruct the three-dimensional density of \Htwo{}. We consider the deprojection as a Bayesian variational inference problem. The posterior distribution of the gas densities allows us to estimate both the mean and uncertainty of the reconstructed density. Unlike most of the previous attempts, we take into account the correlations of gas on a variety of scales which allows curing some of the well-known pathologies, like fingers-of-god effects. Both gas flow models that we adopt incorporate a Galactic bar which induces radial motions in the inner few kiloparsecs and thus offers spectral resolution towards the Galactic centre. We compare our gas maps with those of earlier studies and characterise their statistical properties, e.g. the radial profile of the average surface mass density. We have made our three-dimensional gas maps and their uncertainties available to the community at \href{https://dx.doi.org/10.5281/zenodo.4405437}{this https URL}.
}


\maketitle

\section{Introduction}

Line surveys of molecular gas are a treasure trove for the study of the properties and dynamics of the interstellar medium (ISM~\citealt{2001RvMP...73.1031F}). While the dominant fraction of molecular gas is molecular hydrogen (\Htwo{}), its line emission is inefficient in the cold and dense phase, due to the large spacing of its energy levels. 
The energy levels of carbon monoxide (CO), on the other hand, are much more closely spaced, rendering it radiatively more efficient in the same environments. To a first approximation, the density of \Htwo{} and CO are linearly related, making CO the preferred proxy for \Htwo{}. In particular the $J=1 \to 0$ emission of $\mathstrut^{12} \text{CO}$ at $115 \, \text{MHz}$ has become the observable of choice in the study of the molecular ISM in the Galaxy. While $\mathstrut^{12}\mathrm{CO}$ is almost always in the optically thick regime, the emission from its isotopologue $\mathstrut^{13}\mathrm{CO}$ is in the optically thin regime, thus complementary (e.g.~\citealt{2014MNRAS.445.4055S}), but still bright enough to allow mapping over large regions of the Galaxy (e.g.~\citealt{2021MNRAS.500.3064S}).

Ever since its observational discovery~\citep{1970ApJ...161L..43W}, the coverage, sensitivity and angular resolution of CO surveys have continuously improved. Such surveys have enabled the study of the molecular ISM on scales as small as individual star-forming clumps of a few solar masses up to the total Galactic \Htwo{} mass of the order of $10^9 M_{\odot}$. The applications of CO surveys are thus wide-ranging. Studies of galactic structure on the largest scales, in particular of features such as spiral arms or nuclear bars are interesting in their own right~\citep{2014A&A...569A.125H}, but also offer invaluable clues for the theory of galaxy evolution~\citep{2019ARA&A..57..511K}. The processes at play in the formation and dissolution of molecular clouds can be investigated through the study of individual clouds, but also the statistical analysis of cloud catalogues (e.g.~\citealt{2017ApJ...834...57M}). As stellar nurseries, regions of dense molecular gas are central in the study of star formation~\citep{2012ARA&A..50..531K}. Due to its superior angular resolution compared to broad band emission in other wavelengths, the correlation between different emission processes relies on precise information on the molecular phases. One concrete example is diffuse emission produced by non-thermal cosmic rays~\citep{2012ApJ...750....3A}. Here, the molecular gas (together with atomic gas) provides the target for high-energy cosmic rays, resulting in the production of non-thermal gamma-ray emission, either through bremsstrahlung or the production and subsequent decay of pions into high-energy gamma-rays and neutrinos. While the study of all such processes in the Galaxy is interesting in its own right it is also important to study and calibrate the correlations with other processes in order to use the observations of CO lines for extragalactic astrophysics (e.g.~\citealt{2018ApJ...860..172S}).

Due to our vantage point in the Galaxy the galactic distribution of CO and \Htwo{} is not readily available from the gas line surveys. However, due to Galactic rotation, different emission points along a line of sight in general posses different relative velocities with respect to the observer. Thus, the emission integrated along a line of sight consists of a spectrum that encodes the distribution of emission with distance. The data products of gas line surveys thus consist of spectra for individual lines of sight, oftentimes provided on a three-dimensional grid in longitude $\ell$, latitude $b$ and velocity with respect to the local standard of rest (LSR), $\varv_{\text{LSR}}$. For a given Galactic rotation curve or more generally given a gas flow model, such spectra can be deprojected \emph{in principle} to find the three-dimensional distribution of CO and \Htwo{}.

Unfortunately, a couple of complications hamper the deprojection of the $\ell{}b\varv$-cubes of gas line surveys into a three-dimensional $xyz$-cube of gas densities:
\begin{itemize}
\item Along a given line of sight and for a given velocity, most gas flow models exhibit two distance solutions inside the solar circle. For the velocity range affected by this effect, it is a priori unclear what fractions of the emission are residing at the near and far distances. This effect is commonly referred to as the near-far-ambiguity.
\item For circular rotation, sight lines close to the Galactic centre (longitude $\ell \simeq 0^{\circ}$) and anti-centre (longitude $\ell \simeq \pm 180^{\circ}$) directions exhibit little to no radial velocity, thus lacking kinematical resolution: All emission piles up around $\varv_{\text{LSR}} = 0 \, \text{km} \, \text{s}^{-1}$ and cannot be deprojected along the line of sight.
\item Finally, peculiar velocities, that is random motions of gas on top of the large-scale gas flow, for instance due to stellar winds, supernova explosions or spiral structures, perturb the smooth mapping of distance to radial velocity. Such perturbations become visible as artefacts in the deprojected gas maps. Oftentimes, the distribution of gas gets smeared out along the line of sight, leading to the famous ``finger-of-god'' effect.
\end{itemize}

\begin{figure}[t]
\includegraphics[scale=1]{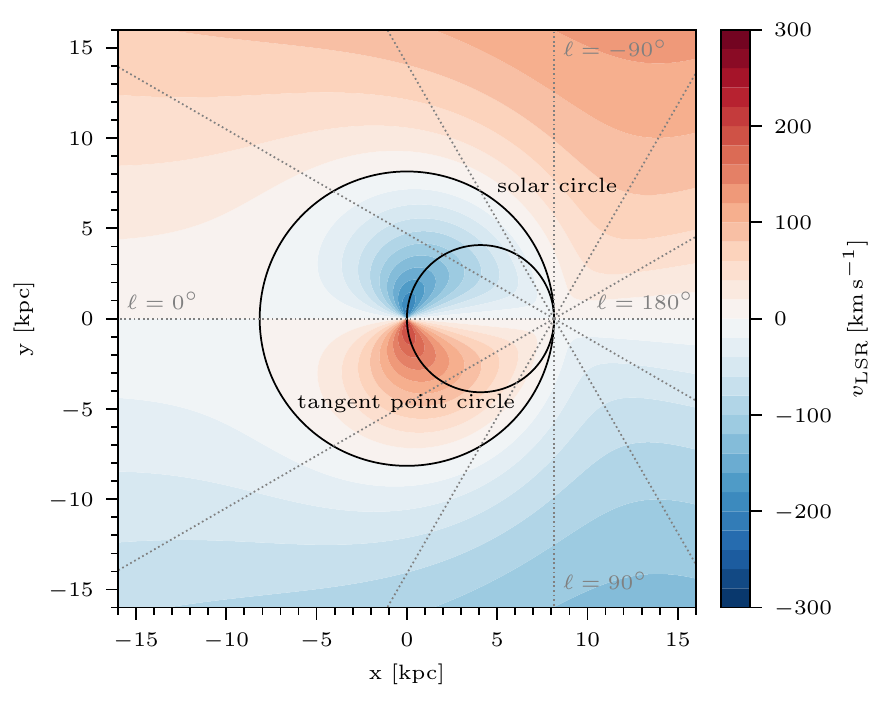}
\caption{Contour plot of the line of sight velocity $\varv_{\text{LSR}}$ for positions $(x,y)$ in the Galactic plane, assuming purely circular motions with speed $V(R) = 220 \, \text{km} \, \text{s}^{-1}$. We also indicate lines of fixed longitude (dotted) and both, the solar circle and the tangent point circles.}
\label{fig:circular}
\end{figure}

We illustrate the first two of these issues for the simple (and unrealistic) example of purely circular rotation. In this case, the radial velocity $\varv_{\text{LSR}}$ along $(\ell, b)$ for emission from the galacto-centric radius $R$ is given by
\begin{equation}
\varv_{\text{LSR}}(R, \ell, b) = \cos b \sin{\ell} \left( \frac{R_{\odot}}{R} V(R) - V_{\odot} \right) \, .
\label{eqn:circular-vLSR}
\end{equation}
Further assuming a flat rotation curve, $V(R) = 220 \, \text{km} \, \text{s}^{-1}$, we show $\varv_{\text{LSR}}$ as a function of position in the Galactic plane ($b = 0^{\circ}$) in Fig.~\ref{fig:circular}. We have assumed the observer to be located at Cartesian coordinates $(x, y) = (R_{\odot}, 0)$ with $R_{\odot} = 8.15 \, \text{kpc}$ and the relation from longitude $\ell$, latitude $b$ and distance along the line of sight $s$ to Cartesian coordinates $x$, $y$ and $z$ is
\begin{align}
    x &= s \cos \ell \cos b - R_{\odot} \, , \label{eqn:x} \\
    y &= s \sin \ell \cos b \, , \label{eqn:y} \\
    z &= s \sin b \, . \label{eqn:z}
\end{align}

The lack of kinematic resolution around the Galactic centre and anti-centre direction is readily visible: Each point along $\ell \simeq 0, \pm 180^{\circ}$ is observed without any Doppler shift, since there is no velocity component along the line of sight. It is also evident that the near-far-ambiguity only affects positions within the solar circle, the circle centred on the Galactic centre with radius $R_{\odot}$. Here, a given velocity $\varv_{\text{LSR}} < 0$ ($\varv_{\text{LSR}} > 0$) corresponds to two solutions for $-90^{\circ} < \ell < 0^{\circ}$ ($0^{\circ} < \ell < 90^{\circ}$). The two solutions are separated by the tangent point circle: a circle of radius $R_{\odot}/2$, centred at $(x, y) = (R_{\odot}/2, 0)$, the locus of points on the line of sight that are tangent to lines of constant $\varv_{\text{LSR}}$. Note that for a given line of sight the tangent point is an extremum of $\varv_{\text{LSR}}$.

There have been a number of previous attempts of deprojecting the results of gas line surveys into three-dimensional gas distributions and each study had to adopt a way for dealing with the issues discussed above. As for the lack of kinematic resolution towards the Galactic centre and anti-centre, \citet{2006PASJ...58..847N} adopted a circular gas flow model, thus ruling out the possibility of directly reconstructing gas near $\ell = 0^{\circ}$ and $\ell = 180^{\circ}$. Instead, they resorted to interpolations between sidelines with small, but finite, radial velocities, at least for this side of the Galactic centre. The region beyond the Galactic centre with little to no kinematical resolution were excluded. \citet{2008ApJ...677..283P} instead adopted the result of a numerical simulation~\citep{2003MNRAS.340..949B} of the gas flow which includes non-circular motions. Towards the Galactic centre and anti-centre this provides several finite kinematic solutions, thus somehow aggravating the distance ambiguity, but also providing some kinematic resolution. Gas could thus be constructed also close to $\ell = 0^{\circ}$ and $\ell = \pm 180^{\circ}$ without interpolation.

As for the near-far-ambiguity, assuming an exponential or Gaussian distribution of gas in the direction perpendicular to the plane of the disk, the projection along the line of sight with a double-valued distance solution leads to a latitude profile that is composed of two Gaussians. Performing fits to the latitude distribution in individual velocity bins, thus allows to determine the relative distribution of gas between the two distance solutions. This method is know as the double Gaussian method and a number of studies have adopted this technique (e.g.~\citealt{1988ApJ...327..139C}). Both \citet{2006PASJ...58..847N} and \citet{2008ApJ...677..283P} have also used the double Gaussian method to break the near-far degeneracy in their deprojection. Specifically, \citet{2008ApJ...677..283P} iteratively deprojected limited velocity ranges, assuming a certain thermal width, until the residual velocity spectra were in agreement with observational noise. Yet, some artefacts are clearly present in their gas maps. \citet{2006PASJ...58..847N}, on the other hand, managed to suppress some potential artefacts, but at the cost of adopting a rather coarse resolution, given the level of detail and small-scale structure available in the CfA CO survey compilation.

An alternative to the deprojection that manages to evade the above mentioned problems to a certain degree is forward modelling. For this, a parametric model for the gas distribution needs to be produced. Recently, \citet{2018ApJ...856...45J} provided such models for atomic and molecular hydrogen and determined the free parameters by fits  to existing survey data. Note that here both the near-far-ambiguity and the lack of kinematic resolution are fixed by assuming a coherent distribution of gas across the offending regions. The authors managed to constrain a large number of parameters by their fit and the properties of the spiral arms bear some resemblence due to other, complementary data sets. However, as the authors admit themselves, not all the gas implied from the survey data can be succesfully deprojected such that the gas maps are to be considered as a lower limit of the real gas maps.

An important additional physical constraint that the gas densities have to fulfil but which is not leveraged by any of the previous studies is the existence of correlations. Such correlations must exist on a range of scales, and are due to a variety of processes. On the largest scales, these are due to the large-scale structure of the disk and the spiral arms, thus ultimately consequences of the formation history and density waves~\citep{2016ARA&A..54..667S}. On smaller scales, the correlations are determined by the turbulent nature of the interstellar medium~\citep{1941DoSSR..30..301K}, affecting both the fluctuation of gas density and velocity as well as magnetic fields. Modelling the three-dimensional gas density as a (Gaussian) random field, these correlation can be parametrised by the power spectrum of gas densities. In fact, there is some support for the hypothesis of a log-normal random field~\cite{1999intu.conf..218N,2001ApJ...546..980O}, thus we will consider the log of gas densities to behave like a Gaussian random field. Such a reconstruction problem for the gas density under the priors of a given correlation structure is most conveniently formulated in a Bayesian framework. We note that with a general enough inference method, the parameters for the power spectrum do not need to be assumed, but can be determined by the inference method together with the gas density. In addition, the Bayesian inference method provides not only an estimate of the gas density, but also quantifies its uncertainty. To our knowledge, none of the previous studies provided such an uncertainty estimate. 

\begin{figure*}
\includegraphics[width=\textwidth]{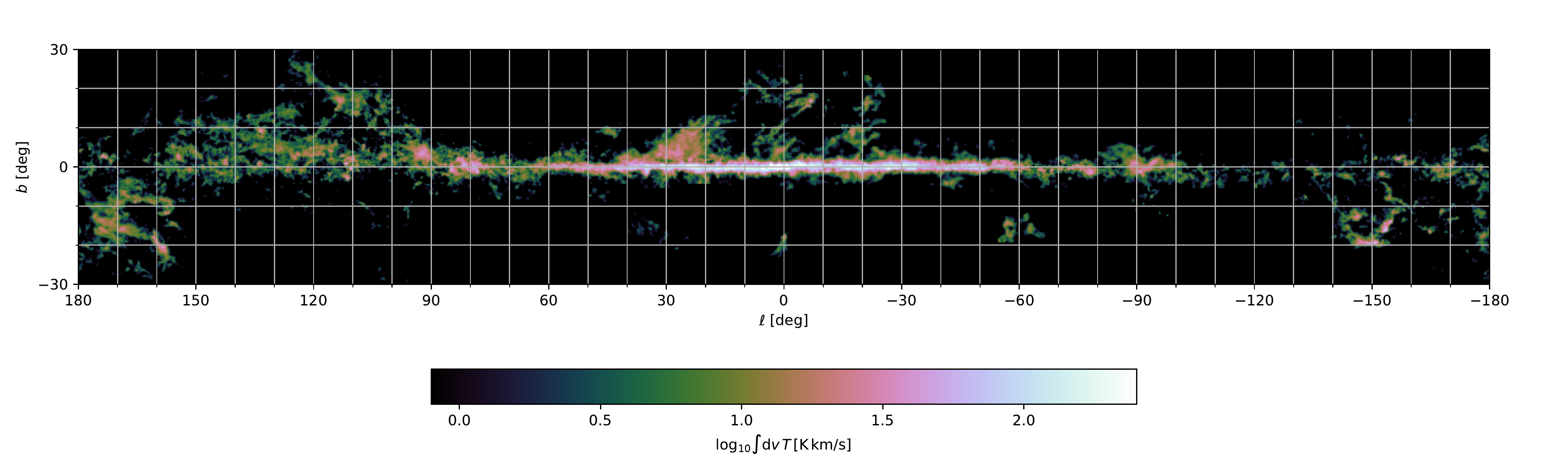} \\
\includegraphics[width=\textwidth]{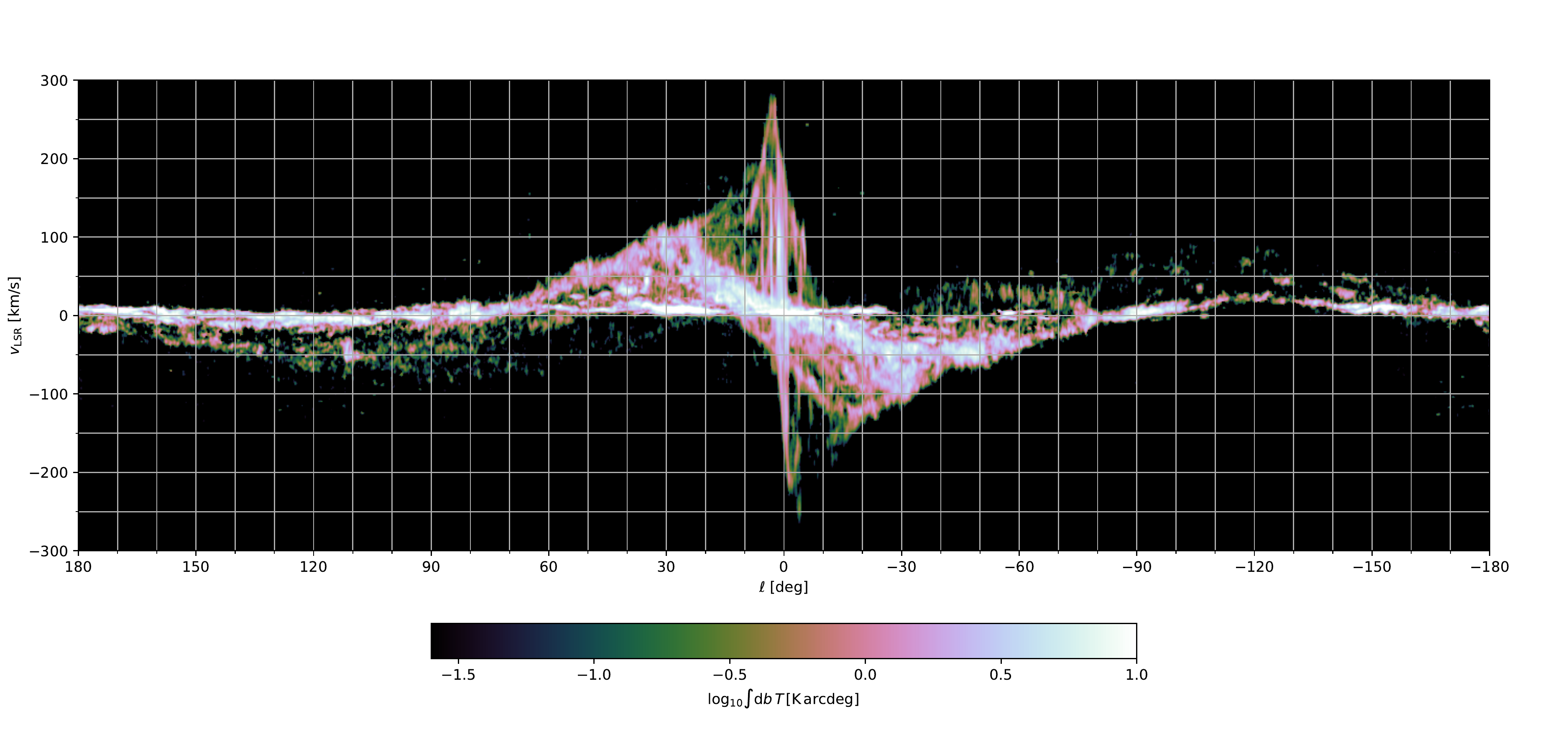}
\caption{\textbf{Top:} Velocity-integrated skymap of Galactic CO emission from the \citet{2001ApJ...547..792D} compilation of surveys, after correcting for the updated local parameters. \textbf{Bottom:} $\ell$-$v$-diagram of Galactic CO emission, integrated over latitudes from $-30^{\circ}$ to $30^{\circ}$, after correcting for the updated local parameters.}
\label{fig:data}
\end{figure*}

One would hope that taking into account the existence of correlations, one would be able to predict gas densities for regions for which data is less constraining, e.g. along $\ell = 0^{\circ}$ and $\ell = \pm 180^{\circ}$. The quality of the deprojected data for these regions will of course be reflected in an increased uncertainty for these regions. Additionally, it can help break the near-far-ambiguity: Due to the coherence on large scales, the two solutions with gas at different distances will not exhibit the same likelihood and the inference algorithm will thus be able to distinguish between those solutions. Finally, taking the presence of observational noise into account in our inference mehtod naturally allows denoising the observations.

The remainder of this paper is organised as follows: In Sec.~\ref{sec:method}, we briefly review the survey data used, introduce the two gas flow models we adopt and present the Bayesian method adopted for the deprojection as well as our data model. Our results are shown in Sec.~\ref{sec:results} in a variety of representations. We compare both our gas maps and supplemental results with those of previous studies. We conclude in Sec.~\ref{sec:Summary} and provide some thoughts on future directions. Details and supplemental information on one of the gas flow models are provided in appendix~\ref{app:gas_flow_model}.

\section{Method}
\label{sec:method}

\subsection{Survey data}

We are using the $\mathstrut^{12} \text{CO}$ ($1 \to 0$) line spectra as compiled by \citet{2001ApJ...547..792D}. This comprises 37 individual surveys that were performed with two 1.2 meter millimeter wave telescopes operated at Columbia University in New York City, NY, at the Centre for Astrophysics in Harvard, MA, and at Cerro Tololo in Chile. Together, these surveys cover the entire Galactic plane in longitude, extending to $\pm 30^{\circ}$ in latitude, thus covering virtually all areas for which significant emission has been reported. The velocity range is from $-319.8$ to $319.8 \, \text{km/s}$ in velocity. Between individual surveys, the angular resolution varies from $1/16$ to $1/2$, as does the rms noise level which we conservatively fix to $0.3 \, \text{K}$ per channel. The survey data were downloaded from the SAO Radio Telescope Data Centre\footnote{\url{https://www.cfa.harvard.edu/rtdc/}}.

Note that before publication the raw survey data had been corrected for the motion of the Earth and the Sun with respect to the LSR. Since the time of the publication of \citet{2001ApJ...547..792D}, better estimates of these relative velocities have become available, however. Therefore we have corrected the survey data by taking into account the updated parameter values as described in Sec.~4.1 of \citet{2018ApJ...856...52W}, adopting the parameter values of a recent, parallax-based determination of the distances to $\sim 200$ masers~\citep{2019ApJ...885..131R}. Specifically, we have coverted the standard cartesian velocity component $(U, V, W) = (10, 15, 7) \, \text{km/s}$ of the original surveys with the values $(U', V', W') = (10.6, 10.7, 7.6) \, \text{km/s}$ of model A5 of \citet{2019ApJ...885..131R}. In Fig.~\ref{fig:data}, we show two projections of the corrected $\ell b \varv$-cube, that is integrated over velocity, the so-called zero-moment map (top panel) and integrated over latitude, the $\ell-\varv$ diagram (bottom panel). In particular the $\ell$-$\varv$ diagram nicely illustrates the coherent structures that will help cure some of the deficiencies of the usual deprojection techniques.

After the deprojection, we will convert from the inferred CO emissivity to \Htwo{} gas density, adopting a linear relation between them. This relation is commonly assumed to exist between the \Htwo{} column density $N_{\mathrm{H}_2}$ and the velocity-integrated CO brightness temperature $W_{\text{CO}} \equiv \int \dd \varv \, T_{\text{b}}$, $N_{\mathrm{H}_2} = X_{\text{CO}} W_{\text{CO}}$, thus establishing the conversion factor $X_{\text{CO}}$. While the relative abundances of \Htwo{} and $\mathrm{CO}$ depend on a number of local factors (e.g.\ density, temperature, metallicity) it is customary to adopt an average value of the order of $10^{20} \, \text{molecules} \, \text{cm}^{-2} \, (\text{K} \, \text{km} \, \text{s}^{-1})^{-1}$. The value $X_{\text{CO}} = 2 \times 10^{20} \, \text{molecules} \, \text{cm}^{-2} \, (\text{K} \, \text{km} \, \text{s}^{-1})^{-1}$ with an uncertainty of $\pm 30 \, \%$ has been recommended~\citep{2013ARA&A..51..207B}. For a recent review of molecular gas surveys, see~\citet{2015ARA&A..53..583H}.

\subsection{Gas flow models}

\begin{figure}
\includegraphics[scale=1]{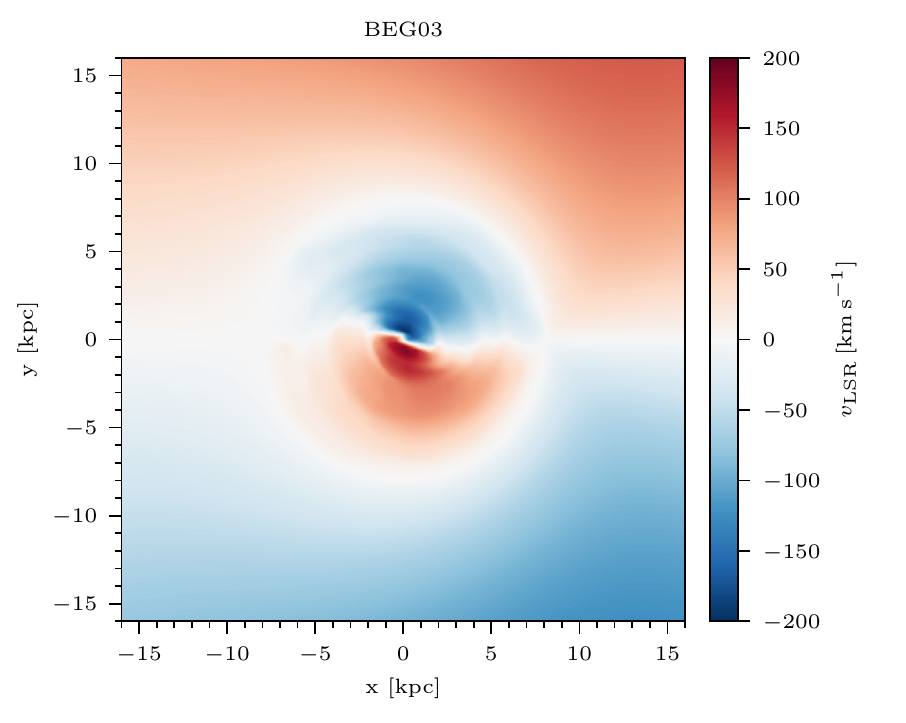}
\includegraphics[scale=1]{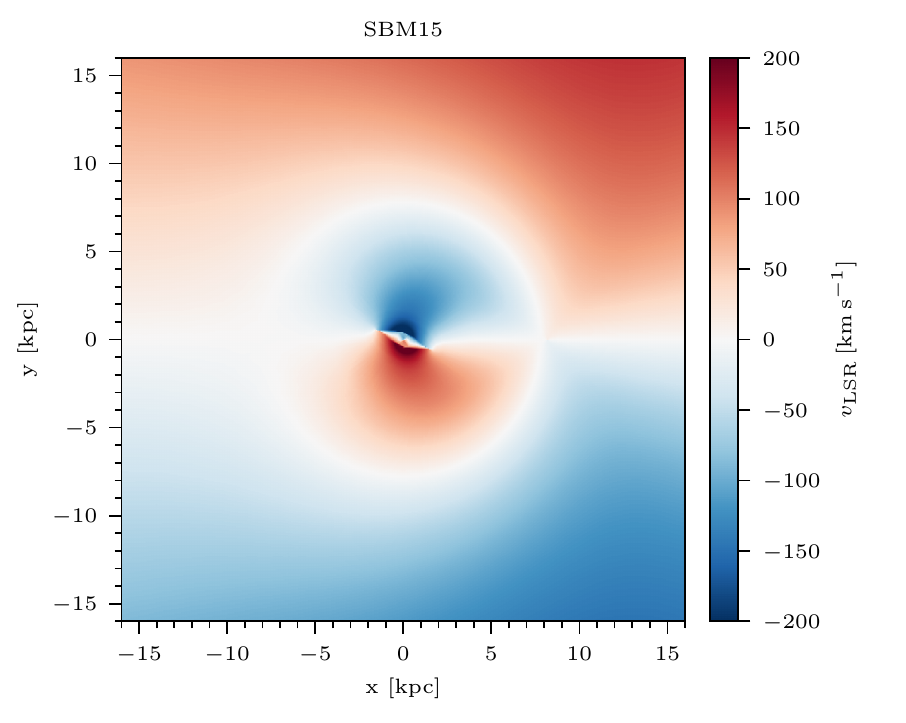}
\includegraphics[scale=1]{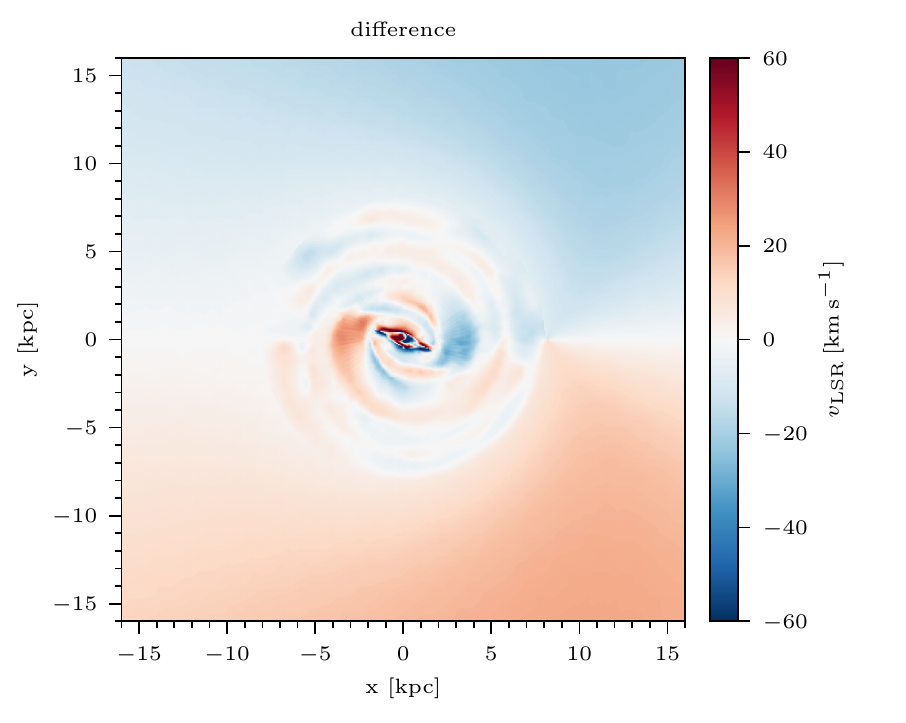}
\caption{Comparison of the two gas flow models that we adopted. \textbf{Top:} $\varv_{\text{LSR}}$ for the BEG03 model. \textbf{Middle:} $\varv_{\text{LSR}}$ for the \citetalias{2015MNRAS.449.2421S} model. \textbf{Bottom:} Difference between $\varv_{\text{LSR}}$ in the BEG03 and the \citetalias{2015MNRAS.449.2421S} model. Note that the color scale is different in the bottom panel.}
\label{fig:vLSR}
\end{figure}

The distribution of CO brightness temperature in $\ell$, $b$ and $\varv$ depends, apart from the three-dimensional gas density field that we wish to reconstruct, on the three-dimensional velocity field which is also unknown. While it might seem that the assumption of purely circular rotation is the most model-independent assumption possible, it is problematic in two ways: practically, as it does not provide any kinematic resolution towards the Galactic centre and anti-centre directions and factually, as the gas flow is known \emph{not} to be purely circular in the inner Galaxy due to the well-established presence of the Galactic bar~\citep{1991ApJ...379..631B} that induces radial motions in the inner few kiloparsec. While the existence of non-circular motions is qualitatively not challenged, the details are everything but certain. In order to estimate the systematic uncertainty due to this, we have adopted two gas flow models.

The first is the result of a smoothed particle hydrodynamics simulation~\citep[][hereafter BEG03]{2003MNRAS.340..949B}. Apart from the gravitational potential of bulge and bar, some assumptions were made about the presence of spiral arms, in particular a four-armed spiral structure was assumed. The resulting $\ell$-$\varv$-diagrams exhibit some resemblance with the CO data, but clearly miss some of the finer structure of the \citet{2001ApJ...547..792D} composite survey. We show their distribution of radial velocities in the top panel of Fig.~\ref{fig:vLSR}. Here, we have extended the gas flow model with a flat rotation curve beyond $8 \, \text{kpc}$, in a similar fashion as \citet{2008ApJ...677..283P}. The assumed presence of the spiral arms is clearly visible in gradients in the velocity field. While newer simulations are available (e.g.~\citealt{2010PASJ...62.1413B,2014MNRAS.444..919P}) we have focussed here on~\citetalias{2003MNRAS.340..949B} to allow for our results to be contrasted with the ones by \citet{2008ApJ...677..283P} who used the same gas flow model.

In addition we have constructed a gas flow model based on a semi-analytical model for gas-carrying orbits in the potential dominated by the Galactic bar~\citep{2015MNRAS.449.2421S}. We provide details of our model in appendix~\ref{app:gas_flow_model}. In the middle panel of Fig.~\ref{fig:vLSR}, we show the resulting radial velocity map as seen by an observer at a distance of $8.15 \, \text{kpc}$ from the Galactic centre~\citep{2019ApJ...885..131R} at an angle of $20^{\circ}$ with respect to the major axis of the bar and moving with the velocity of the LSR. We also show the difference between the velocity fields of~\citetalias{2003MNRAS.340..949B} and the \citetalias{2015MNRAS.449.2421S} model in the bottom panel of Fig.~\ref{fig:vLSR}. Three major differences are evident: First, while the overall agreement inside the solar circle is good, the agreement in the inner $2 \, \text{kpc}$ is less so. Second, the perturbations due to the spiral structure present in the model of \citetalias{2003MNRAS.340..949B} are clearly absent in our \citetalias{2015MNRAS.449.2421S} model. The difference is mostly of the order of $\pm 10 \, \text{km/s}$, but can be as large as $\pm 30 \, \text{km/s}$ in limited regions. Finally, outside the solar circle the difference in the adopted rotation curves is marked and can lead to differences as large as $\pm 30 \, \text{km/s}$. Note however that there is little molecular gas at the relevant distances.

\subsection{Bayesian inference}

In deprojecting gas line surveys, we follow two connected goals: First, we want to reconstruct the three-dimensional gas density under the constraint that the density possesses a certain spatial correlation structure. While we can be sure that such correlations exist, the details are not well-constrained and ideally we would hope for the data to constrain the correlation structure itself. Second, we would also like to obtain an estimate of the uncertainties. Both these goals are most directly achieved by adopting a Bayesian framework.

We consider the deprojection of the three-dimensional gas density from line survey data as a high-dimensional Bayesian inference problem, that is we are seeking the posterior distribution of the gas density for the given the survey data. According to Bayes' theorem~\citep{1763RSPT...53..370B} the posterior is proportional to the likelihood, that is the probability for the observed brightness temperature given the gas density, times the prior, that is the probability of the gas density. We could constrain ourselves to finding the maximum of the posterior, however, the maximum can be uninteresting if the posterior is multi-modal or has degenerate directions. In addition, we would need to estimate the uncertainty separately, e.g.\ adopting the Laplace approximation~\citep{Laplace:1774zz} that equates the covariance at the maximum posterior position with the inverse Hessian. Instead, it is advantageous to keep track of the uncertainty while exploring the posterior.

Monte Carlo Markov Chain (MCMC) methods~\citep{Hastings:1970aa} do exactly that and they can approximate arbitrary posteriors given large enough sample sizes. However, with growing dimensionality, they become computationally expensive and for the large dimensionality of the current problem, prohibitively so. Variational inference~\citep{2016arXiv160100670B} instead approximates the posterior with a parametric distribution, for instance a multivariate Gaussian. The parameters of the parametric distribution can be determined if the ``distance'' between the approximate distribution and the true posterior can be estimated, for instance through the Kullback-Leibler divergence~\citep{1968its..book.....K}. For a multi-variate Gaussian, this would in principle involve the inversion of the large covariance matrix which is again computationally prohibitive. Instead, it has been suggested~\citep{2019arXiv190111033K} to approximate the covariance with the inverse Fisher information metric, a method known as Metric Gaussian Variational Inference. This method has recently been applied to problems ranging from reconstruction of the three-dimensional dust density in the Galaxy from reddening data~\citep{2019A&A...631A..32L} to radio interferometry~\citep{2019A&A...627A.134A}.

In practice, this is implemented as an iterative scheme, alternating between estimating the covariance at the current mean and updating the mean for the current estimate of the covariance. Specifically, adopting standardisation of the parameters, the computation of the (inverse) Fisher information metric requires the Jacobian of the standardisation map. This can be obtained either analytically or numerically by automatic differentiation. The mean is estimated by minimising the Kullback-Leibler divergence with respect to the mean. Note  that this does not require explicitly computing the covariance matrix, which would entail the inversion of the Fisher information metric. Instead, the Kullback-Leibler divergence can be estimated stochastically, that is by drawing samples from a Gaussian with said covariance, which can be implemented with implicit operators. The application of the covariance in drawing the samples constitutes a linear system that can be solved by using a conjugate gradient algorithm.

\subsubsection{Data model}

The relation between the CO emissivity $\varepsilon(x, y, z)$ that we want to reconstruct and the brightness temperature $T(\ell,b,v)$ from the \citet{2001ApJ...547..792D} survey is given by a linear map $R$ from signal space $(x,y,z)$ to data space $(\ell, b, \varv)$,
\begin{equation}
R [ \varepsilon ](\ell, b, \varv) = \int_0^{\infty} \dd s \, \varepsilon (\vec{r}) \, \delta(\varv - \varv_{\text{LSR}}(\vec{r})) \Big|_{\vec{r} = \vec{r}(\ell, b, s)} \, ,
\label{eqn:response_map}
\end{equation}
with $\vec{r}(\ell, b, s)$ specified in eqs.~\eqref{eqn:x} to~\eqref{eqn:z} above. Setting $T = R [ \varepsilon ]$ would result in a deterministic likelihood $p(T | \varepsilon)$,
\begin{equation}
p(T_{\ell b \varv} | \varepsilon_{xyz} ) = \delta \left( T_{\ell b \varv} - R [ \varepsilon_{xyz} ] \right) \, ,
\end{equation}
but we have to take into account the presence of additive noise $n$, thus altering out data model to
\begin{equation}
T_{\ell b \varv} = R [ \varepsilon_{xyz} ] + n_{\ell b \varv} \, .
\label{eqn:data_model}
\end{equation}
We assume this noise to be Gaussian distributed, $n \sim p(n) = \mathcal{G}(n, N)$, with covariance $N$ that is diagonal in harmonic space, $\tilde{N} = 2 \pi \sigma_n^2 \delta(\vec{k} - \vec{k}')$, that is white noise. Given the properties of the individual surveys combined in \citet{2001ApJ...547..792D} we fix $\sigma_n = 0.3 \, \text{K}$. We marginalise over the noise, thus obtaining a Gaussian likelihood,
\begin{align}
\tilde{p}(T_{\ell b \varv} | \varepsilon_{xyz}) &= \int \dd{n} \, p(T_{\ell b \varv} | \varepsilon_{xyz}, n_{\ell b \varv} ) p(n_{\ell b \varv}) \\
&=  \int \dd{n} \, \delta \left( T_{\ell b \varv} - R [ \varepsilon_{xyz} ] - n_{\ell b \varv} \right) \mathcal{G}(n, N) \, .
\end{align}
Finally, we have to take into account the fact that the measured $\varv$ can differ from the actual radial velocity $\hat{\varv}$ of the emitting gas, e.g.\ due to thermal line width or turbulence. We take the difference $(\varv - \hat{\varv})$ to be normal distributed with variance $\sigma_{\varv}^2$ and fix $\sigma = 5 \, \text{km} \, \text{s}^{-1}$. In principle, this requires another marginalisation, but we approximate this through a smearing of the linear map $R$ instead,
\begin{align}
& p(T_{\ell b \varv} | \varepsilon_{xyz}) \\ &\equiv \int \dd \hat{\varv} \int \dd{n} \, \delta \left( T_{\ell b \hat{\varv}} - R [ \varepsilon_{xyz} ] - n_{\ell b \varv} \right) \mathcal{G}(n, N) \mathcal{G}(\varv - \hat{\varv}, \sigma_{\varv}^2) \\
&\simeq \int \dd{n} \, \delta \left( T_{\ell b \varv} - R' [ \varepsilon_{xyz} ] - n_{\ell b \varv} \right) \mathcal{G}(n, N) \\
&= \mathcal{G} \left( T_{\ell b \varv} - R' [ \varepsilon_{xyz} ], N \right) \, ,
\end{align}
where
\begin{align}
R' [ \varepsilon_{xyz} ] &\equiv \int \dd \hat{\varv} \, \mathcal{G}(\varv - \hat{\varv}, \sigma_{\varv}^2) R [ \varepsilon_{xyz} ] \\
&= \int_0^{\infty} \dd s \, \varepsilon(\vec{r}) \, \mathcal{G}(\varv - \varv_{\text{LSR}}(\vec{r}), \sigma_{\varv}^2) \Big|_{\vec{r} = \vec{r}(\ell, b, s)} \, .
\end{align}

To be able to define the posterior distribution, we still need to specify the signal prior. We model the CO emissivity $\varepsilon(\vec{r})$ as a log-normal distributed random field, that is $s(\vec{r}) \equiv \ln (\varepsilon(\vec{r}) / \varepsilon_0)$ is normal distributed. Under the assumption that $s$ is statistically homogeneous, i.e.\ the two-point correlation in configuration space, $\langle s(\vec{r}) s(\vec{r}') \rangle$, is a function of the distance |$\vec{r} - \vec{r}'|$ only, the two-point correlation in the harmonic domain becomes ``diagonal'', that is $\langle \tilde{s}(\vec{k}) \tilde{s}(\vec{k}') \rangle = 2 \pi P(k) \delta(\vec{k} - \vec{k}')$ where $P(\vec{k})$ is the power spectrum. We further assume that the power spectrum $P(\vec{k})$ is isotropic, $P(\vec{k})$ = $P(k)$.

Instead of assuming a form for the power spectrum $P(k)$, we would like to determine it during the reconstruction as well. Following \citet{2019A&A...631A..32L}, we therefore adopt a statistical model for the power spectrum, with a Gaussian distributed normalisation $y$, a Gaussian distributed power law index $m$ and add a Gaussian random field in $\log(k)$,
\begin{align}
\sqrt{P(k)} = \exp & \left[ (\mu_y + \sigma_y \phi_y) + (\mu_m + \sigma_m \phi_m) \log(k) \right. \\
& \left. + \mathcal{F}^{-1} \!\! \left\{ \frac{a}{1 + t^2/t_0^2} \tau(t) \right\} \right] \! .
\end{align}
Here, $\phi_y$ and $\phi_m$ are random variables and $\tau(t)$ a random field (in $\log(k)$) that are encoding the power spectrum and are reconstructed at the same time as the signal itself. $\mathcal{F}^{-1}$ represents the inverse Fourier transform from the variable $t$ that is the conjugate of $\log(k)$. The parameters, $\mu_y$, $\sigma_y$, $\mu_m$, $\sigma_m$, $a$ and $t_0$ are meta-parameters and we have fixed them to the following values: $\mu_y = -13$, $\sigma_y = 0.1$, $\mu_m = -4$, $\sigma_m = 0.1$, $a = 1$ and $t_0 = 0.1$. We stress that the above representation of $P(k)$ is flexible enough to closely approximate the true underlying power spectrum for $k$ where the data is constraining enough; where it is not, the shape is interpolated or extrapolated.

Finally, assuming $s$ to be a homogeneous random field must fail when the deviations from this assumption for the real gas density become too strong. This is certainly the case when considering the confinement of molecular gas to the Galactic plane. In order to not be biased in the reconstruction, we have scaled the log-normal signal field $\varepsilon_{xyz}$ with an exponential profile in $z$, $\exp[ - |z|/z_h ]$. We have adjusted $z_h$ to values such that the signal field $\varepsilon_{xyz}$ averaged over different portions of the Galactic disk does not show strong gradients in the $z$-direction and found $z_h = 40 \, \text{pc}$ to give satisfactory results. We have also tested a Gaussian profile $\exp[ - z^2/(2 \sigma_z^2) ]$, but found the results to be virtually unchanged.

\subsubsection{Details on the implementation}

As we are performing our computations on a computer, the signal field is not a continuous field of gas densities, but rather a discretised version thereof. We adopt a Cartesian grid with the $x$, $y$ and $z$-coordinates indexed with $\alpha,\beta,\gamma$, that is $\varepsilon_{\alpha \beta \gamma} = \varepsilon(x_\alpha, y_\beta, z_\gamma)$. Specifically, we have considered a $512 \times 512 \times 16$ Cartesian grid stretching from $-16 \, \text{kpc}$ to $16 \, \text{kpc}$ in the $x$- and $y$-directions and from $-0.5 \, \text{kpc}$ to $0.5 \, \text{kpc}$ in the $z$-direction, thus achieving a spatial resolution of $1/16 \, \text{kpc} = 62.5 \, \text{pc}$.

Given that the data are the spectra from a binned survey, they are already discretised, that is $T_{i j k} = T(\ell_i, b_j, \varv_k)$. We have degraded the data from their native resolution on a $2880 \times 481 \times 493$ grid to a $1440 \times 241 \times 247$ grid. The discrete version of the linear map of eq.~\eqref{eqn:response_map} is
\begin{equation}
R_{ijk}^{\alpha \beta \gamma} \!\! = \!\! \int \! \dd \ell \, \dd b \, \dd \varv \, \dd s \, \frac{\varepsilon_{\alpha\beta\gamma}}{\Delta\ell \Delta b \Delta \varv}  \theta(\text{if $\vec{r}$ in ${\alpha\beta\gamma}$} ) \delta(\varv \! - \! \varv_{\text{LSR}}(\vec{r})) \Big|_{\vec{r} = \vec{r}_{ijk}} \! .
\end{equation}
Our data model, eq.~\eqref{eqn:data_model}, can thus be represented by a multiplication with the sparse matrix $R_{ijk}^{\alpha \beta \gamma}$.

For a specific implementation of the Gaussian variational inference, we have made use of the \texttt{nifty5} package\footnote{\url{https://gitlab.mpcdf.mpg.de/ift/nifty}}.

\section{Results and Discussion}
\label{sec:results}

\begin{figure*}[!th]
\includegraphics[width=\textwidth]{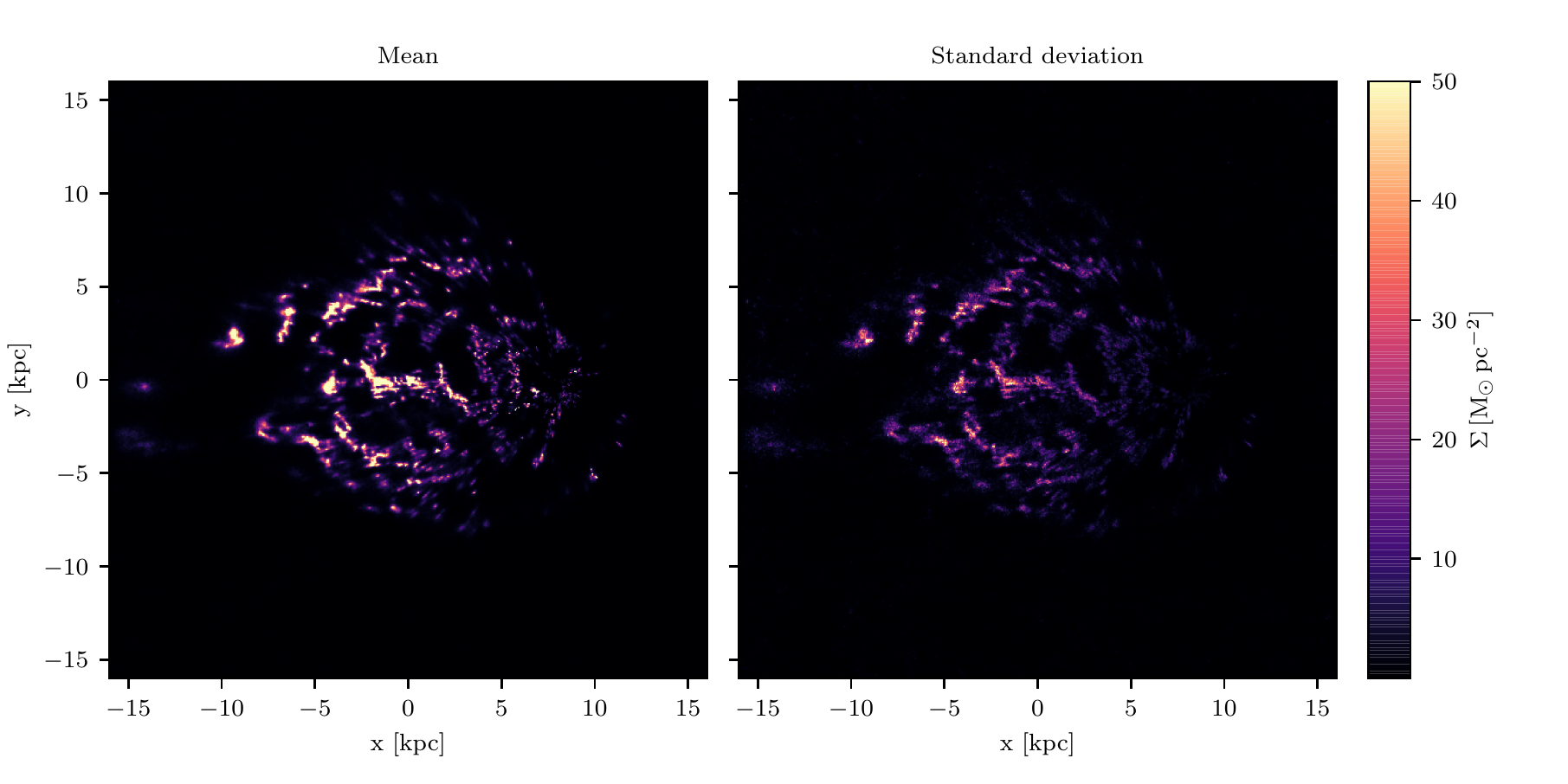} \\
\includegraphics[width=\textwidth]{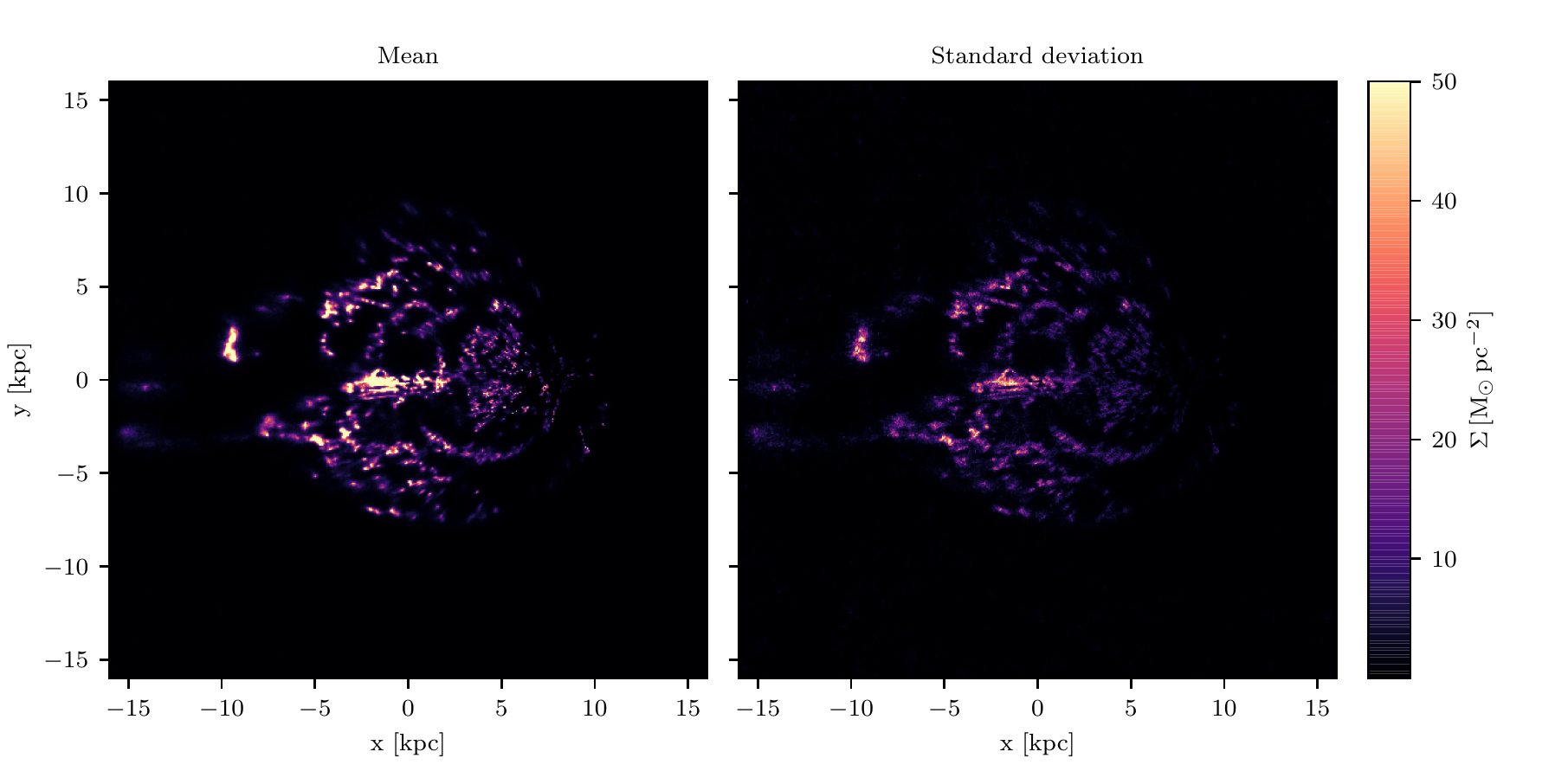}
\caption{Two-dimensional projection of reconstructed three-dimensional maps of molecular hydrogen. \textbf{Top left:} Mean gas surface density $\Sigma$ for the \citetalias{2003MNRAS.340..949B} gas flow model. \textbf{Top right:} Standard deviation of the gas surface density $\Sigma$  for the \citetalias{2003MNRAS.340..949B} gas flow model. \textbf{Bottom left:} Mean gas surface density $\Sigma$ for the \citetalias{2015MNRAS.449.2421S} model. \textbf{Bottom right:} Standard deviation of the gas surface density $\Sigma$ for the \citetalias{2015MNRAS.449.2421S} model.}
\label{fig:mean_uncert}
\end{figure*}

The main result of our analysis are the 3D gas maps obtained from the two gas flow models and the corresponding uncertainties. Our gas maps provide the highest resolution three-dimensional deprojections of CO gas line surveys to date with a number of robust and well-localised emission regions and should prove useful in the study of Galactic structure and diffuse emission. We make our maps available to the community\footnotemark.

\footnotetext{\url{http://dx.doi.org/10.5281/zenodo.4405437}}

In Fig.~\ref{fig:mean_uncert} we show the projection of the \Htwo{} density onto the Galactic plane and its standard deviation, both for the BEG2003 gas flow model (top panels) and the \citetalias{2015MNRAS.449.2421S} gas flow model (bottom panels). For either gas flow model, the survey data have been successfully deprojected into localised clusters of emission. The total gass mass reconstructed is $1.1 \times 10^9 M_{\odot}$ for the \citetalias{2003MNRAS.340..949B} model and $1.6 \times 10^9 M_{\odot}$ for the \citetalias{2015MNRAS.449.2421S} model.

Some elongated structures, spurs or spiral arm segments are immediately visible. Some of the structures are more easily visible for the \citetalias{2003MNRAS.340..949B} model than for the \citetalias{2015MNRAS.449.2421S} model. For instance, the two vertical spurs stretching from $(x, y) = (4, -1) \, \text{kpc}$ to $(x, y) = (4, 3) \, \text{kpc}$ and from $(x, y) = (6, -1) \, \text{kpc}$ to $(x, y) = (6, 3) \, \text{kpc}$, respectively, are easily identified for the \citetalias{2003MNRAS.340..949B} model (top left panel of Fig.~\ref{fig:mean_uncert}), but blend into a more extended emission for the \citetalias{2015MNRAS.449.2421S} model. Revisiting Fig.~\ref{fig:vLSR} and in particular its bottom panel, it is clear that such structures are oftentimes linked to local extrema in the radial velocity field. For instance the two spurs discussed above coincide with maxima of the velocity field of the \citetalias{2003MNRAS.340..949B} model, but are absent in the \citetalias{2015MNRAS.449.2421S} model. These are due to the spiral arms which have been put in by hand in the case of the \citetalias{2003MNRAS.340..949B} model, but not in the \citetalias{2015MNRAS.449.2421S} model. We can also identify a spur that coincides with the tangent point circle, stretching from $(x, y) = (2, -3) \, \text{kpc}$ to $(x, y) = (6, -4) \, \text{kpc}$. This is likely an artefact pointing to too narrow a velocity range in the \citetalias{2015MNRAS.449.2421S} model.

\begin{figure*}[p]
\centering
\includegraphics[scale=1]{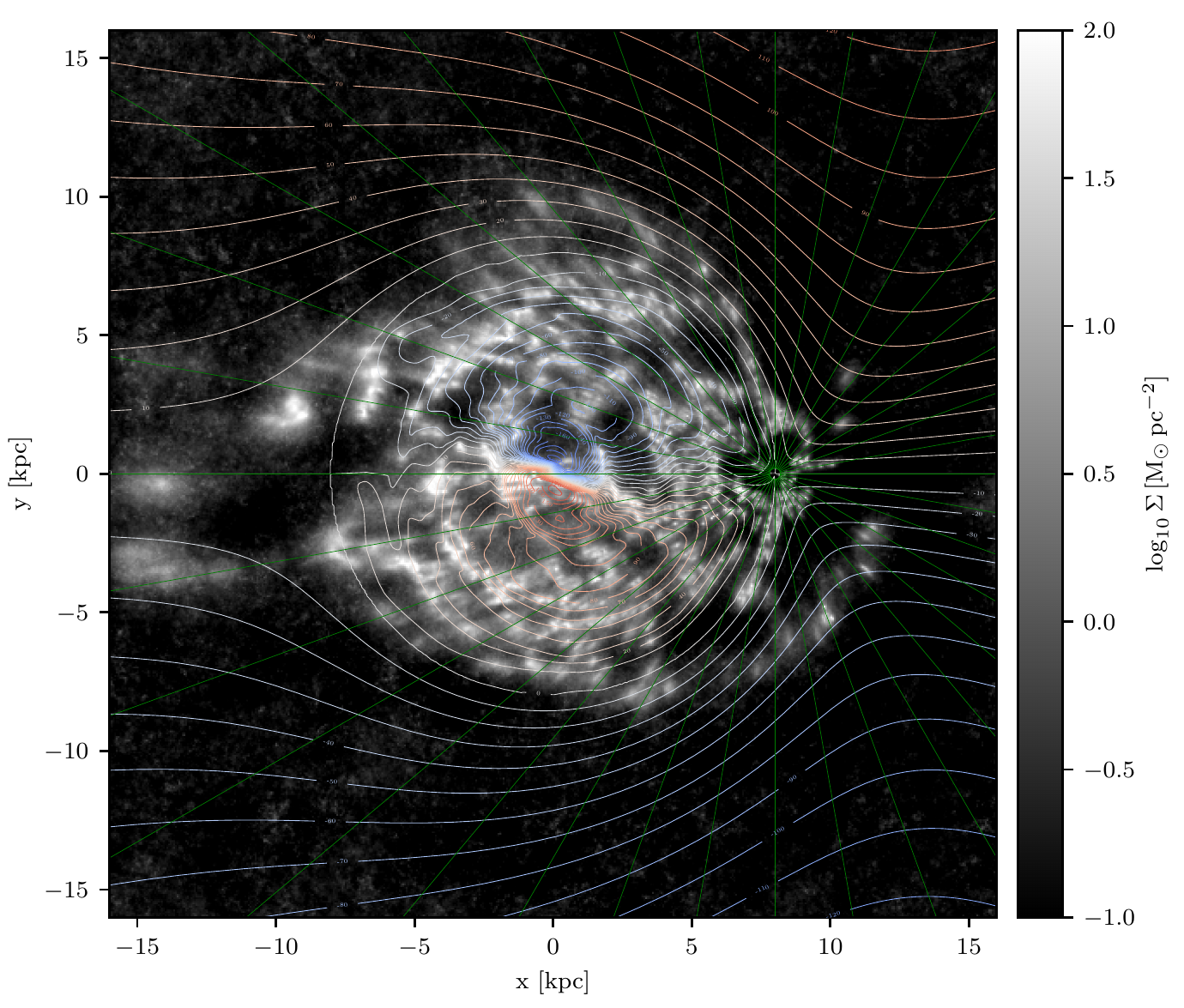} \\
\includegraphics[scale=1]{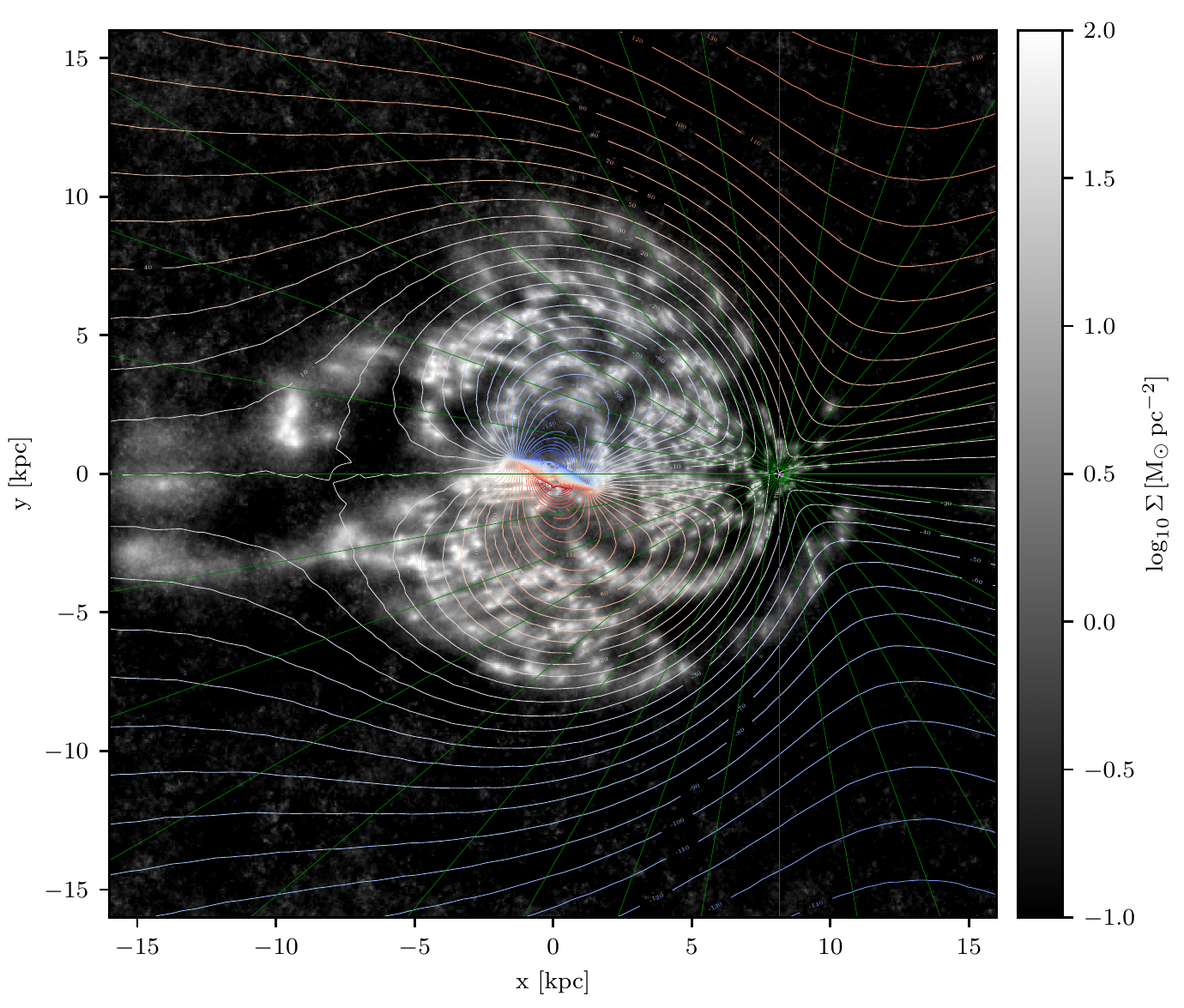}
\caption{Projected mean gas density $\Sigma$ on a logarithmic colour scale, overlaying both contours in $\varv_{\text{LSR}}$ and a grid in longitude. \textbf{Top:} For the \citetalias{2003MNRAS.340..949B} gas flow model. \textbf{Bottom:} For the \citetalias{2015MNRAS.449.2421S} model.}
\label{fig:details1}
\end{figure*}

We note that the gas density comprises a rather large dynamical range as expected for a log-normal density. Some of the fainter features are therefore difficult to identify on a linear scale. In order to highlight them, we show the mean of the posterior, again projected onto the Galactic plane on a logarithmic colour scale in Fig.~\ref{fig:details1}. We have also overlaid a longitude grid and the contours of the respective radial velocities. This should allow to more easily identify certain features with the corresponding features in the $\ell-\varv$-diagram of Fig.~\ref{fig:data}.

\begin{figure*}[p]
\centering
\includegraphics[scale=1]{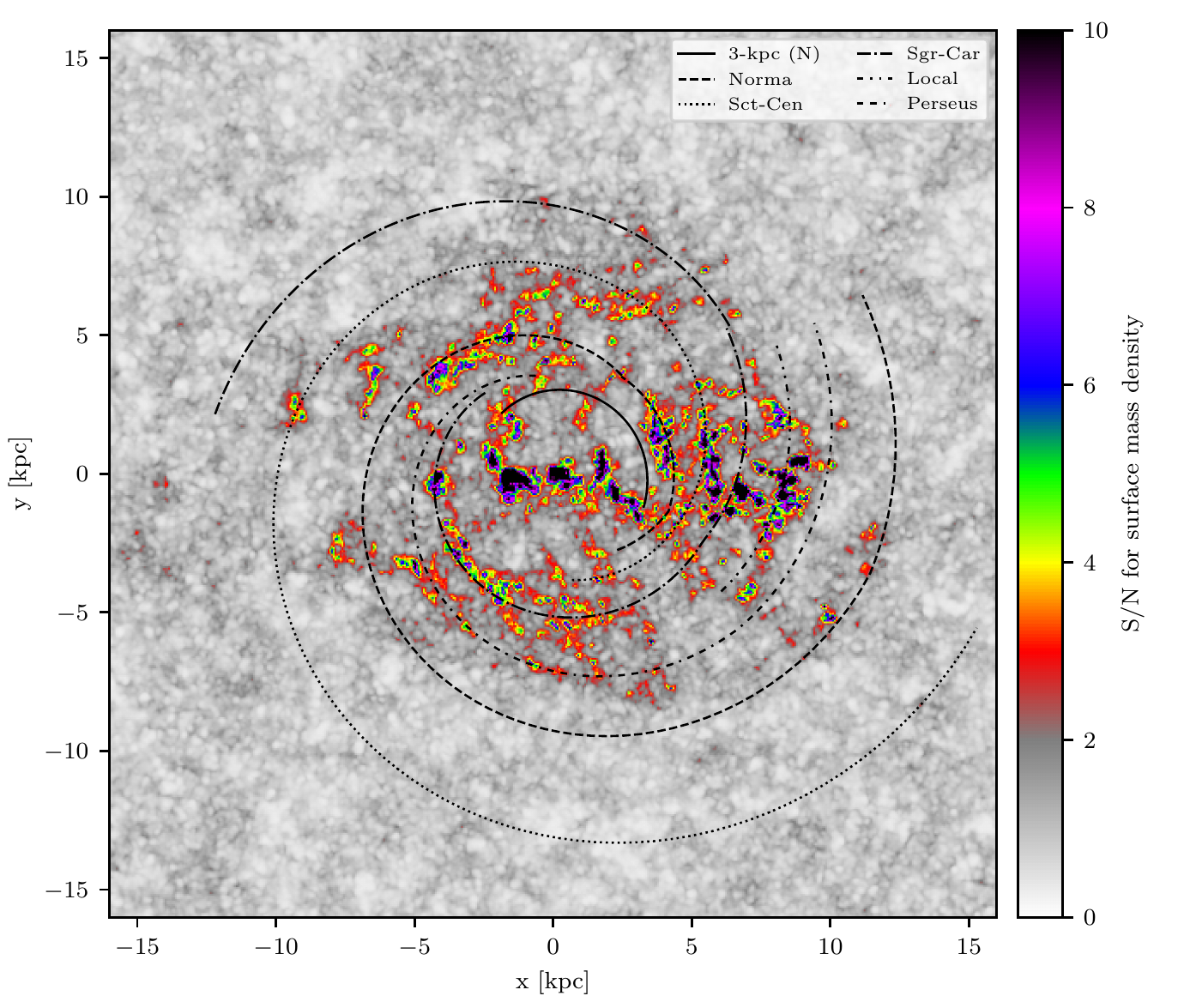} \\
\includegraphics[scale=1]{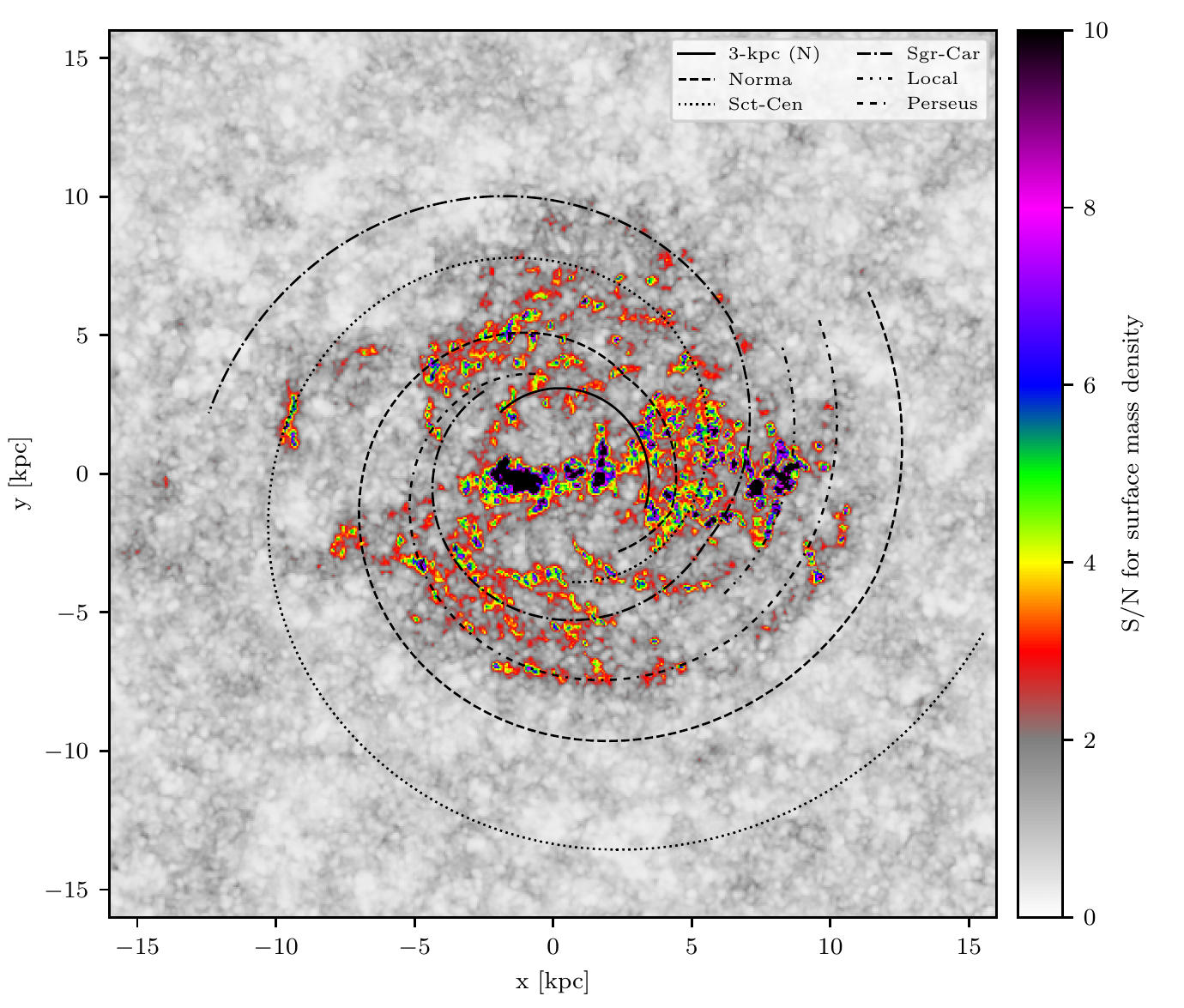}
\caption{Signal-to-noise ratio S/N of the mean gas density, with the spiral arms of \citet{2019ApJ...885..131R} overlaid. \textbf{Top:} For the \citetalias{2003MNRAS.340..949B} gas flow model. \textbf{Bottom:} For the \citetalias{2015MNRAS.449.2421S} model.}
\label{fig:details2}
\end{figure*}

\begin{figure*}[t]
\includegraphics[scale=1]{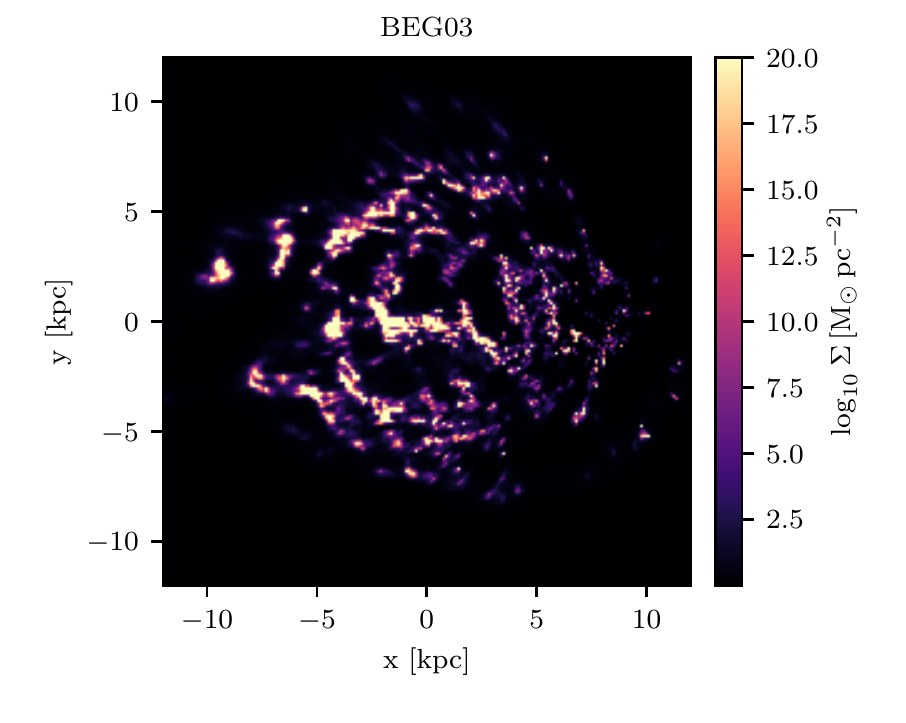} \includegraphics[scale=1]{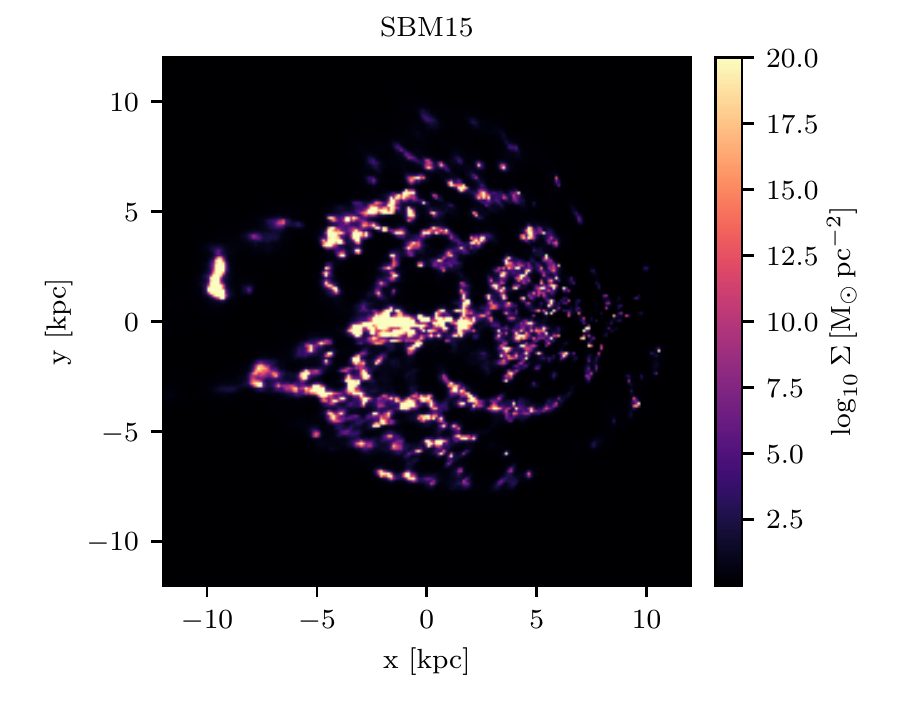} \\
\includegraphics[scale=1]{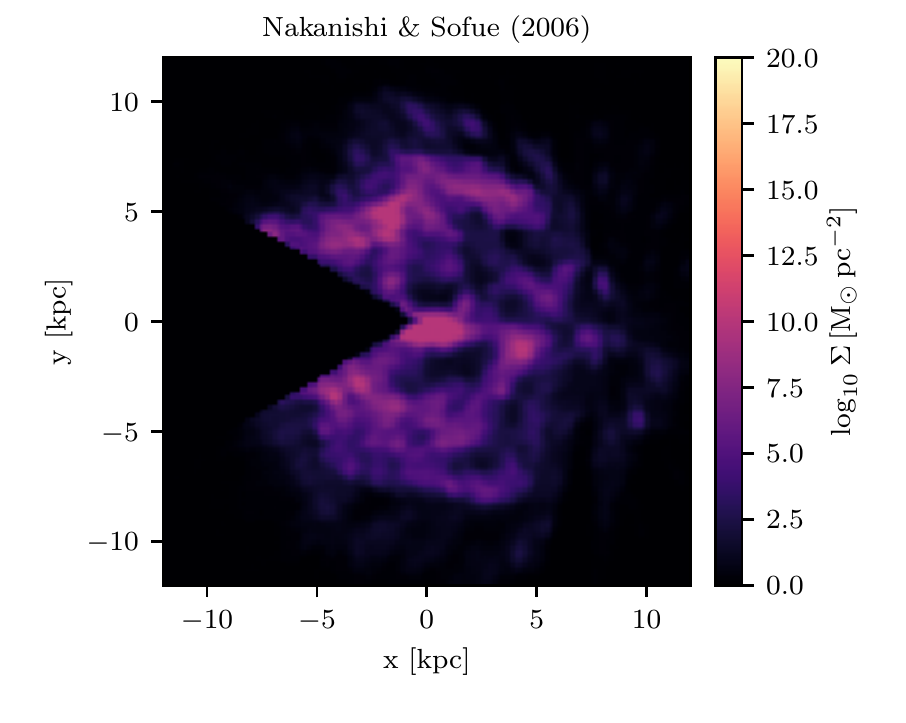} \includegraphics[scale=1]{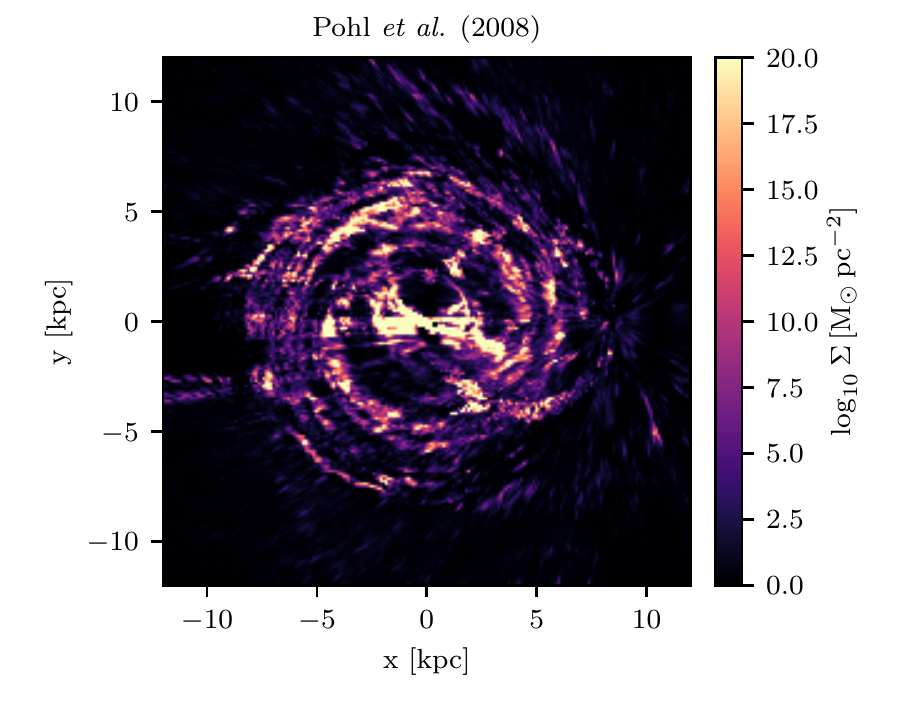} \\
\caption{Comparison of both our gas surface density construtions (top left and top right) with those of \citet{2006PASJ...58..847N} (bottom left) and \citet{2008ApJ...677..283P} (bottom right).}
\label{fig:comparison}
\end{figure*}

However, looking at the mean of the posterior alone can be misleading as some of the localised features also have a rather large uncertainty. Unlike the previous deprojections, however, we now have a means of judging the validity of certain features by comparing the mean $\mu$ of the posterior with its uncertainty $\sigma$. To this end, we define a signal-to-noise ratio (S/N) as $\mu / \sigma$. We show S/N in Fig.~\ref{fig:details2}, again for the \citetalias{2003MNRAS.340..949B} model in the top panel and in the bottom panel for the \citetalias{2015MNRAS.449.2421S} model. One can clearly identify localised emission with a S/N ratio of 3 or higher. In Fig.~\ref{fig:details2}, we have also overlaid the spiral arms, as determined from fits to a set of $\sim 200$ masers~\cite{2019ApJ...885..131R}. (See their Tbl.~2 for the fitted spiral parameters.) Many of the local emission features obtained for either gas flow model can be easily associated with a spiral arm: for the \citetalias{2003MNRAS.340..949B} model for all spiral arms, but most impressively for the Norma, Sagittarius-Carina, Local and Perseus arms. We comment on a couple of noteworthy differences and similarities between the significant features obtained for the \citetalias{2003MNRAS.340..949B} and the \citetalias{2015MNRAS.449.2421S} models:
\begin{itemize}
\item The gas density in the \citetalias{2015MNRAS.449.2421S} model is generally more scattered and does not cluster in regions as large as the emission in the \citetalias{2003MNRAS.340..949B} model. This is again due to the presence of local extrema in the radial velocity field in the \citetalias{2003MNRAS.340..949B} model which boost the clustering. Such local extrema are all but absent in the \citetalias{2015MNRAS.449.2421S} model and hence the gas density is less clustered.
\item Yet, some of the spiral arms are obvious also for the \citetalias{2015MNRAS.449.2421S} model, e.g. the segments along the Scutum-Centaurus and the Sagittarius-Carina arms for galacto-centric azimuths $\varphi$ between $\sim 200^{\circ}$ and $\sim 280^{\circ}$. Other examples are the segments along the Norma arm ($90^{\circ} \lesssim \varphi \lesssim 150^{\circ}$), the local arm ($330^{\circ} \lesssim \varphi \lesssim 0^{\circ}$).
\item Some emission, in particular beyond the solar circle, is placed at different distances in the \citetalias{2003MNRAS.340..949B} and \citetalias{2015MNRAS.449.2421S} models due to the different rotation curves adopted here. Given the rather small velocity gradient, this easily translates into differences of the order of a kiloparsec and thus affects the association with spiral arms. One example is emission around $\ell \sim 110^{\circ}$ and with $\varv_{\text{LSR}}$ between $-60$ and $-50 \, \text{km}/\text{s}$, see Fig.~\ref{fig:data}, bottom panel. With the \citetalias{2003MNRAS.340..949B} model, this emission is located around $(x, y) = (10, -6) \, \text{kpc}$. With the \citetalias{2015MNRAS.449.2421S} gas flow, this instead ends up at $(x, y) = (9.5, -5) \, \text{kpc}$. In the former case, an association with the Norma arm suggests itself, in the latter case, the association with the Perseus arm is more likely.
\end{itemize}

In Fig.~\ref{fig:comparison}, we revisit the mass surface densities obtained with either gas flow model (top panels) and compare them with the deprojections of \citet{2006PASJ...58..847N} (bottom left) and \citet{2008ApJ...677..283P} (bottom right). Note that all gas densities are shown with the same dynamical range, however, the mass surface density of \citet{2006PASJ...58..847N} is significantly smoother. Some similarities are apparent between the map of \citet{2008ApJ...677..283P} and our maps. This is even more so the case for our map that is based on the \citetalias{2003MNRAS.340..949B} model, the same gas flow model adopted by \citet{2008ApJ...677..283P}. However, there are also some differences:
\begin{itemize}
\item Spiral arms are more homogeneous in \citet{2008ApJ...677..283P}. While in our reconstruction, the width is oftentimes varying along the segments, in the gas density of \citet{2008ApJ...677..283P}, some ring segments appear to have constant width. This might be an artefact of the particular algorithmic reconstruction chosen there.
\item An artefact, that is affecting both the \citet{2008ApJ...677..283P} map and our \citetalias{2015MNRAS.449.2421S} map is the emission concentrated along an arc of the tangent point circle (see discussion above).
\item The peak in the gas density reconstructed for both our gas flow models near $(x, y) = (-9, 2) \, \text{kpc}$ is absent in the \citet{2008ApJ...677..283P} map, possibly due to the chosen suppression at large galacto-centric radii.
\item Another class of artefact, that is apparent in the \citet{2008ApJ...677..283P} maps are the features elongated along the lines of sight, the so called ``fingers of god'', especially in regions of low gas densities. These are almost completely absent in our reconstructions.
\item Finally, \citet{2008ApJ...677..283P} show some gas density in regions where we have found (almost) none, in particular beyond the solar circle. Unlike the structures solidly identified with spiral arms of the \citetalias{2003MNRAS.340..949B} model, some of these might actually be statistically not significant. This is to be compared with emission seen for our reconstruction in Fig.~\ref{fig:details1} near $(x, y) = (-14, -1) \, \text{kpc}$ and $(x, y) = (-14, -4) \, \text{kpc}$ which is not statistically significant, see Fig.~\ref{fig:details2}.
\end{itemize}

\begin{figure}[htb]
\includegraphics[scale=1]{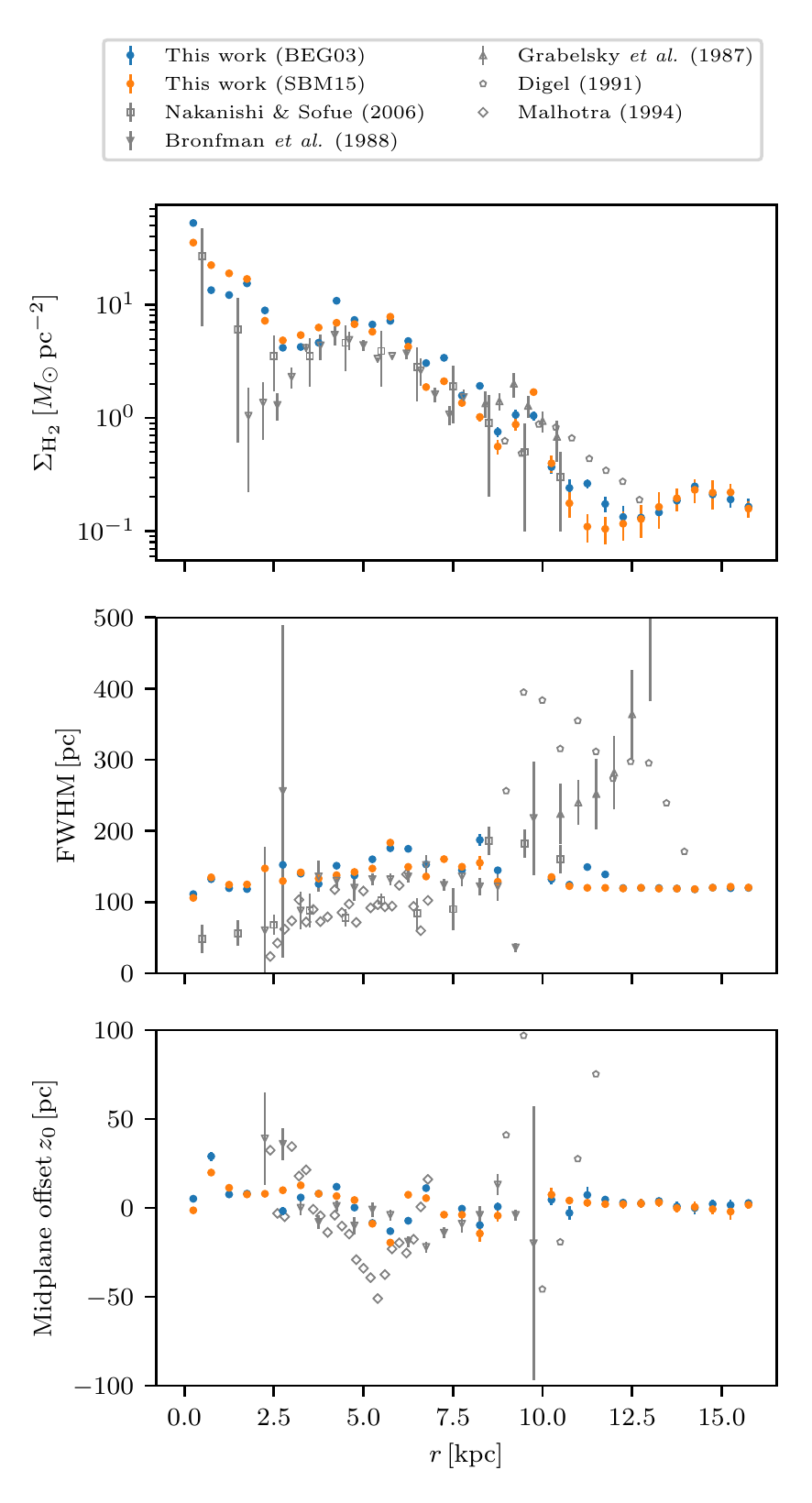}
\caption{Radial profiles of the surface mass density $\Sigma_{\mathrm{H}_2}$ (top panel), FWHM of the $z$-profile (middle panel) and midplane offset (bottom panel). We compare the results for both our gas flow models (\citetalias{2003MNRAS.340..949B} and \citetalias{2015MNRAS.449.2421S}) with the results of earlier studies: \citet{1987ApJ...315..122G}, \citet{1988ApJ...324..248B} and \citet{2006PASJ...58..847N}.}
\label{fig:profiles}
\end{figure}

We conclude our discussion by looking at some properties of the derived distributions in galacto-centric radius. In Fig.~\ref{fig:profiles}, we have plotted the surface mass density averaged in galacto-centric rings $R_i \leq R < R_{i+1}$ where $R_i = i \Delta R$ with $\Delta R = 0.5 \, \text{kpc}$. Close to the Galactic centre, the average gas density is a few tens of $M_{\odot} \, \text{pc}^{-2}$ and subsequently decreases towards $r \simeq 3 \, \text{kpc}$. Beyond, there is a local maximum at $r \simeq 5 \, \text{kpc}$, the well-established molecular ring (or possibly a convergence of spiral arms~\cite{2012MNRAS.421.2940D}). Here, we find the average gas density to be $\sim 1 \, M_{\odot} \, \text{pc}^{-2}$. Beyond, the gas density decreases further. While it appears to reach a plateau of $2 \times 10^{-2} M_{\odot}$ around $12 \, \text{kpc}$, we recall that we have not detected much significant emission beyond $r \simeq 10 \, \text{kpc}$.

We have fitted Gaussian profiles $g_i(z) \propto \exp[-(z-z_{0,i})^2/2 \sigma_i^2]$ with means $z_{0,i}$ and standard deviations $\sigma_i$ to the gas density distribution in the same galacto-centric rings $i$ as before by minimising the $\chi^2$ over all rings,
\begin{equation}
\chi^2 \equiv \sum_i \sum_{\gamma} \frac{ \langle n_{\Htwo{},\alpha\beta\gamma} \rangle_{\alpha\beta,i} - g_i(z_\gamma) }{\tilde{\sigma}_i^2} \, .
\end{equation}
Here, $\langle \mathellipsis \rangle_{\alpha\beta,i}$ denotes the averaging over all grid points $\alpha\beta$ in the $i$-th ring. We have fixed the uncertainty $\tilde{\sigma}_i$ to $0.15$ of the maximum of $\langle n_{\Htwo{},\alpha\beta\gamma} \rangle_{\alpha\beta,i}$ for that particular ring $i$. For most rings, the Gaussian profile is a fair approximation, $\chi^2 < 1$. (Of course, different $z$ will be correlated, hence $\chi^2$ is usually $< 1$.) For other rings, however, the average profile in $z$ is only poorly described by a single Gaussian (see also~\citealt{1985ApJ...297..751D}). We have thus removed these bins. We stress that the ensemble of samples from the posterior consists of full 3D distributions. Also uncertainties and correlations contained, so allow for any statistical analysis beyond the axisymmetric quantities and simple error estimates.

We show the FWHM, that is $2 \sqrt{2 \log 2} \sigma$, a measure of the vertical extent of the gas distribution, as a function of galacto-centric radius in the middle panel of Fig.~\ref{fig:profiles} and compare to a number of previous estimates. We note that our reconstructed gas densities seems to be more extended vertically, at least in the inner few kiloparsecs. Note that \citet{2008ApJ...677..283P} does not determine this parameter but instead uses it as an input for the analysis.

In the bottom panel of Fig.~\ref{fig:profiles}, we show the dependence of the midplane offset $z_0$ on the galacto-centric radius and again compare to a number of previous estimates. While the general trend with galacto-centric radius is the same as that seen in previous studies, our error bars are consistently smaller. We stress that the ensemble of samples from the posterior consists of full 3D distributions. Also uncertainties and correlations contained, thus allowing for any statistical analysis beyond the axisymmetric quantities and simple error estimates.

\section{Summary}
\label{sec:Summary}

We have presented a new deprojection of the CO line survey of~\citet{2001ApJ...547..792D} with unprecedented spatial resolution of $62.5 \, \text{pc}$. This is based on a Gaussian variational Bayesian inference which allows exploring the posterior distribution of this high-dimensional inference problem. While we have assumed correlations in configuration space to exist, we have not assumed any particular power spectrum, but determined the power spectrum during the reconstruction. We have considered two gas flow models, that both take into account the presence of the Galactic bar, one based on a simulation of the gas flow in a predetermined potential~\citep[]{2003MNRAS.340..949B}[BEG03], the other based on a a model for gas carrying orbits in the bar potential~\citep[]{2015MNRAS.449.2421S}[SBM15].

Our results are the three-dimensional distribution of molecular gas, assuming a fixed $X_{\text{CO}}$ factor of $2 \times 10^{20} \, \text{molecules} \, \text{cm}^{-2} \, (\text{K} \, \text{km} \, \text{s}^{-1})^{-1}$. We have made our mean gas maps and their uncertainty available to the community\footnotemark[\value{footnote}]. We have shown the mean and the standard deviation of the gas density projected onto the Galactic plane and compared with previous studies. Unlike those earlier studies, we have the capacity to distinguish between statistically significant structures and noise artefacts. We have found that some of the most prominent structures are influenced by the assumed spiral structure in the \citetalias{2003MNRAS.340..949B} gas flow model, but that significant coherent structures, some of which align with spiral arms (as for instance defined by parallax measurements of masers) are present in the \citetalias{2015MNRAS.449.2421S} model as well. We have projected some radial profiles out of the 3D gas distribution and found the results of previous studies largely confirmed.

In the future, a couple of extensions of the current analysis are noteworthy. Having successfully applied our methodology to molecular line surveys to determine the \Htwo{} density it would be interesting to apply it to surveys of atomic hydrogen, like the recently completed HI4PI survey~\citep{2016A&A...594A.116H}. While the HI distribution is known to show less clustering than molecular hydrogen, we are optimistic that the advantages of our approach carry over. In addition, a couple of shortcomings of the present study are due to our ignorance of the gas flow. If the velocity field could be determined at the same time as the gas density, such deficiencies could be remedied. We can envisage two ways to help regularise this underdetermined problem: Data from the above mentioned parallax measurements~\citep{2019ApJ...885..131R} could be included, thus providing at least a couple of loci to which the velocity field could be anchored. In addition, physical correlation between gas densities and flow velocities could link the reconstruction of both fields, of course, this would require additional model inputs.

\bibliographystyle{aa}
\bibliography{gift}

\begin{appendix}
\section{Semi-analytical gas flow model}
\label{app:gas_flow_model}

Short of running our own hydrodynamical simulations of gas flow in a barred potential, we employ a semi-analytical approximation to the gas flow. It has been hypothesised by \citet{1991MNRAS.252..210B} that gas in the potential of a rotating bar is drifting slowly towards the centre, moving on orbits from two classes of closed orbits, so-called $x_1$ and $x_2$ orbits. The transition from the outer $x_1$ orbits to the inner $x_2$ orbits is taking place via a shock structure that forms along a critical, cusped orbit. Beyond, the $x_1$ orbits are self-intersecting, but the gas is drifting on $x_2$ orbits instead. This picture was observed in early simulations~\citep{1992MNRAS.259..345A}, albeit with limited resolution and a somewhat different potential than assumed by \citet{1991MNRAS.252..210B}, and more recently confirmed for the same potential and with a high resolution simulation~\citetalias{2015MNRAS.449.2421S}. Here, we adopt the potential of~\citetalias{2015MNRAS.449.2421S}, but slightly adjust their parameters.

Specifically, we adopt a combination of a triaxial potential for the bar,
\begin{equation}
\rho(r') = \frac{\rho_0}{4 \pi a^3} \frac{1}{(r' / a)^{\alpha} (1 + r' / a)^{\beta - \alpha}}
\end{equation}
with $r' = \sqrt{x^2 + (y/a)^2 + (z/a)^2}$ and a razor-thin disk potential generated by the surface mass density
\begin{equation}
\Sigma(R) = \Sigma_0 \mathrm{e}^{-R / R_d} \, ,
\end{equation}
$R$ denoting the cylindrical galacto-centric radius. We adopt the following parameters: $\rho_0 = 0.69 M_{\odot} \, \text{pc}^{-3}$, $a=1.8 \, \text{kpc}$, $\alpha = 1.75$, $\beta = 3.5$, $\Sigma_0 = 1.3 \times 10^3 M_{\odot} \, \text{pc}^{-2}$ and $R = 4.5 \, \text{kpc}$. We assume a pattern speed of $\Omega_p = 63 \, \text{km} \, \text{s}^{-1} \, \text{kpc}^{-1}$ for the rigidly rotating bar potential.

\begin{figure}[t]
\includegraphics[width=\columnwidth]{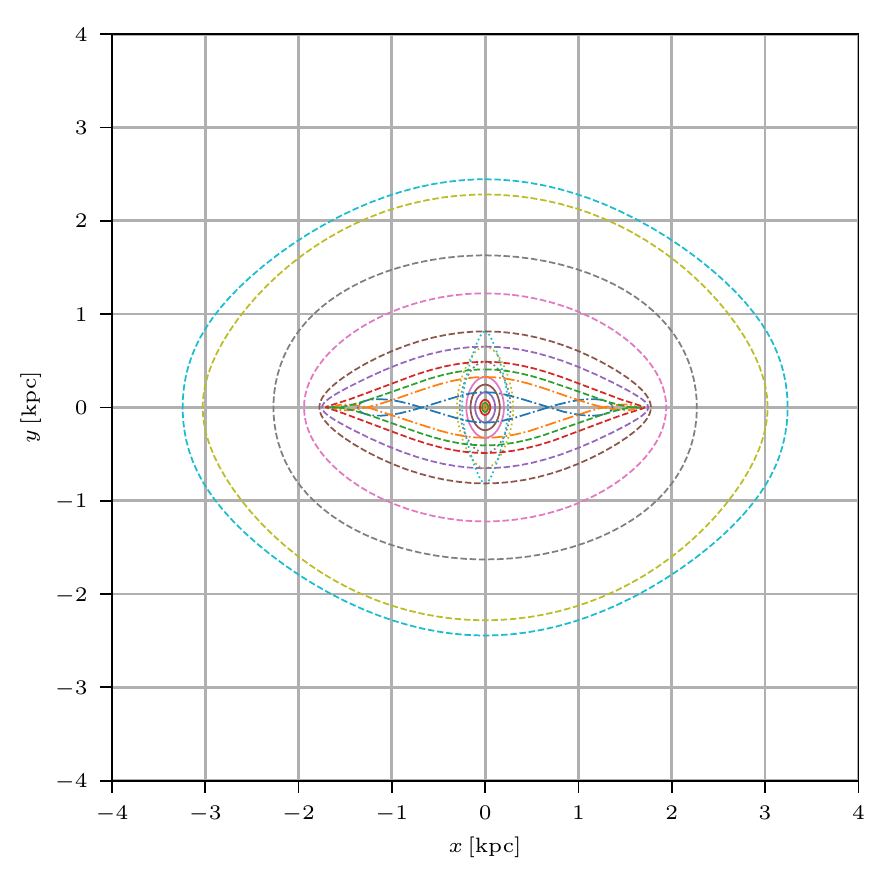}
\caption{Some examples for $x_1$ and $x_2$ orbits found by solving for closed orbits in the rotating bar potential. The $x_1$ orbits are marked by dashed and dot-dashed lines and the $x_2$ orbits are marked by solid and dotted lines. Gas is assumed to be moving on only a subset of those, that is on the orbits marked by a solid or dashed line. Orbits beyond the largest shown orbit are assumed to be circular.}
\label{fig:orbits}
\end{figure}

We solve the equations of motion for test particles in the combined potential using the \texttt{galpy} package\footnote{\url{http://github.com/jobovy/galpy}}~\citep{2015ApJS..216...29B}. We find closed $x_1$ and $x_2$ orbits in the range of galacto-centric radius $\leq 4 \, \text{kpc}$. Some examples are shown in Fig.~\ref{fig:orbits}.

We have interpolated the velocity field between the points on the $x_1$ and $x_2$ orbits for galacto-centric radii $R \leq 4 \, \text{kpc}$. While this works well for the regions inside the largest populated $x_2$ orbit and outside the smallest populated $x_1$ orbit, in between those orbits, the gas veolicites are somewhat overestimated, see Fig.~5 of~\citetalias{2015MNRAS.449.2421S}. We have therefore added a couple of nodes in the interpolation of the radial gas velocities to reach better agreement with Fig.~5 of~\citetalias{2015MNRAS.449.2421S}.

Beyond $4 \, \text{kpc}$, we have assumed gas to move on circular orbits and for the velocity to follow the ``universal rotation curve'' of~\citet{1996MNRAS.281...27P} with the updated parameters of~\citet{2019ApJ...885..131R}. A small supression of $5 \, \%$ was necessary to make the rotation curve connect smoothly with the interpolation at $R = 4 \, \text{kpc}$.

In the middle panel of Fig.~\ref{fig:vLSR}, we have shown the radial velocity resulting from this semi-analytical gas flow model. Even though we have adopted slightly different parameters than \citetalias{2015MNRAS.449.2421S}, we refer to this model as the \citetalias{2015MNRAS.449.2421S} model in the main text.

\end{appendix}

\end{document}